\documentclass[twocolumn,showpacs,preprintnumbers,amsmath,amssymb,prl]{revtex4}

\usepackage[dvips]{graphicx}
\usepackage{dcolumn}
\usepackage{bm}

\def \imagi {\mathrm{i}}
\def \vc {v_{\mathrm{c}}}
\def \vxc {v_{\mathrm{xc}}}
\def \vhx {v_{\mathrm{hx}}}
\def \vhxc {v_{\mathrm{hxc}}}
\def \Ic {I_{\mathrm{c}}}
\def \IcA {I_{\mathrm{c}}^{\mathrm{A}}}
\def \gxc {g_{\mathrm{xc}}}
\def \gx {g_{\mathrm{x}}}
\def \gc {g_{\mathrm{c}}}
\def \gcA {g_{\mathrm{c}}^{\mathrm{A}}}
\def \nA {n^{\mathrm{A}}}
\def \rhoA {\rho^{\mathrm{A}}}

\def \epsce {\epsilon_{\mathrm{ce}}}
\def \epsee {\epsilon_{\mathrm{ee}}}

\begin{document}
\preprint{APS}

\title{Adiabatic Approximation of the Correlation Function in the Density-Functional Treatment of Ionization Processes}

\author{F.~Wilken}
\author{D.~Bauer}
\affiliation{Max-Planck-Institut f\"ur Kernphysik, Postfach 103980,
69029 Heidelberg, Germany}

\date{\today}

\begin{abstract}
The ionization of a one-dimensional model Helium atom in short laser pulses using time-dependent density functional theory is investigated. We calculate ionization probabilities as a function of laser intensity by approximating the correlation function of the system adiabatically with an explicit dependence on the fractional number of bound electrons. For the correlation potential we take the derivative discontinuity at integer numbers of bound electrons explicitly into account. This approach reproduces ionization probabilities from the solution of the time-dependent Schr\"odinger equation, in particular the so-called ``knee'' due to nonsequential ionization.
\end{abstract}

\pacs{31.15.Ew, 32.80.Rm}
\maketitle

Experimental double ionization yields for Helium atoms in laser fields of $\lambda \! = \! 614\, \mathrm{nm}$ \cite{FittinghoffBolton1992} and $\lambda \! = \! 780\, \mathrm{nm}$ \cite{WalkerSheehy1994}  show a ``knee''- structure as a function of intensity, i.e.,  double ionization is enhanced over the expected yields for sequential ionization by several orders of magnitude and was therefore named nonsequential double ionization (NSDI).

Theoretically, the problem was addressed successfully using the strong-field approximation (see, e.g., \cite{Becker19992005} and references therein) and classical methods \cite{Fu2001Ho2005}. The widely accepted mechanism behind NSDI relies on the rescattering of the first electron with its parent ion, collisionally ionizing (or exciting) the second electron. 

However, investigations using time-dependent density functional theory (TDDFT) \cite{RungeGross1984} failed to reproduce the characteristic ``knee'' of the double ionization probability \cite{Bauer1997,LappasLeeuwen1998,BauerCeccherini2001}. This failure is especially important given the fact that TDDFT offers the possibility to treat the strong-field dynamics of atomic and molecular systems from first principles in a numerically tractable way, whereas the solution of the Schr\"odinger equation in full dimensionality \cite{ParkerMoore2000} is limited to two active electrons. Other methods used to tackle the NSDI problem from first principles are R-Matrix Floquet theory \cite{FengvanderHart2003} or multi-configurational time-dependent Hartree-Fock \cite{DahlenLeeuwen2001,Kitzler2004Caillat2005}, which, however, have not yet reproduced NSDI-''knees'' either. 

Two reasons for the failure of TDDFT in the context of NSDI were identified \cite{PetersilkaGross1999}: The approximation of the exchange-correlation functional $\vxc$ and the approximation of the functionals for the calculation of the ionization probabilities. Initial attempts to overcome these problems were not successful \cite{PetersilkaGross1999}. Recent progress in approximating $\vxc$ has been made by 
including the derivative discontinuity at integer numbers of bound electrons
\cite{LeinKuemmel2005,MundtKuemmel2005,deWijnKuemmelLein2006II}.
In this work we therefore focus our attention on the open problem of constructing functionals for the calculation of the ionization probabilities. We show that by using a rather simple adiabatic approximation of the correlation function, the ionization yields for a model He-atom in few-cycle laser pulses (for which we have the exact time-dependent Schr\"odinger solutions at hand) are remarkably well reproduced.

We investigate Helium in linearly polarized 
$N \! = \! 3$-cycle
laser pulses of duration $T \! = \! N\,2\,\pi / \omega$ with a $\mathrm{sin}^{2}$-pulse envelope in dipole approximation, i.e., the vector potential is of the form $A\left( t \right) = \hat{A} \: \mathrm{sin}^{2} \left( \frac {\omega}{2 N} \, t \right) \, \sin \left( \omega \, t \right)$ for $0\! \leq \! t \! \leq \! T$. 
Two
different pulses with frequencies used in experimental work \cite{FittinghoffBolton1992,WalkerSheehy1994} are chosen: $\omega=0.058$ 
and $\omega=0.074$
 (atomic units are used throughout this work unless otherwise indicated).
The linear polarization of the laser pulses allows to model Helium by a one-dimensional atom with soft-core potentials for the Coulomb interactions. It is known that the essential features of the nonsequential double ionization process are described well by this model \cite{Bauer1997,LappasLeeuwen1998,LeinGross2000,DahlenLeeuwen2001,LeinKuemmel2005,deWijnKuemmelLein2006II}.
For different effective peak intensities $I=I ( \hat A \, )$ of the laser pulses we solve the time-dependent Schr\"odinger equation (TDSE) $\imagi \, \partial_{t} \, \psi  = \widehat{H} \, \psi$ with the Hamilton operator 
\begin{equation}
\widehat{H} = \sum_{k=1,2} \left( - \frac{1}{2} \, \partial^{2}_{x_{k}} + V \left( x_{k}, t \right) \right) + W \left( x_{1}-x_{2} \right ) \, ,
\label{Schroedinger Hamilton Operator}
\end{equation}
where $\widehat{H}=\widehat{H}\left (x_{1},x_{2},t \right )$, the external potential $V \left( x, t \right)
= \imagi \, A \left( t \right ) \partial_{x} - 2 / \sqrt{x^{2}+\epsce}$ (the $A^{2}$-term has been transformed away) and the electron-electron interaction potential $W \left( x \right ) = 1/ \sqrt{x^{2} +\epsee}$. The solution is the two-electron wavefunction $\psi \left( x_{1}, x_{2},t \right)$ in a spin singlet state.
Therefore we have just one Kohn-Sham orbital fulfilling $\imagi \, \partial_{t} \, \phi  = \widehat{H}_{\mathrm{KS}} \, \phi$ with $\phi = \phi \left( x,t \right)$ and
$\widehat{H}_{\mathrm{KS}} \left( x,t \right) = - \left( 1/2 \right) \, \partial^{2}_{x} + V \left( x,t \right)+ \vhxc \left( x, t \right)$.
The Hartree-exchange part $\vhx= v_{\mathrm{h}} + v_{\mathrm{x}}$ of the potential $\vhxc = v_{\mathrm{h}} + v_{\mathrm{x}} + \vc$ is known exactly for the Helium atom, $v_{\mathrm{h}} \left( x, t \right) =  \int \mathrm{d}x' \, 2 \, \vert \phi \left( x',t \right) \vert^{2} \, W \left( x-x' \right)$ and $v_{\mathrm{x}} \left( x, t \right) = - \frac{1}{2} \, v_{\mathrm{h}} \left( x, t \right)$.
A simple model for the correlation potential which takes into account the derivative discontinuity at integer numbers of bound electrons is the recently proposed 
\cite{LeinKuemmel2005} 
\begin{equation}
\vhxc^{\mathrm{LK05}}\left( x, t \right)=\frac{\textstyle N_{0}/N \! \left( t \right)}{1+\mathrm{exp} \left[ C \left(N_{0}/N \! \left( t \right)-2\right) \right]} \, \vhx \left( x, t \right) \, ,
\label{Formula vhxcLK05}
\end{equation}
where $N \! \left( t \right) \! = \! \int_{-a}^{+a} \mathrm{d}x \, 2\, \vert \phi \left( x,t\right) \vert^{2} $ is the number of bound electrons and $N_{0} \! = \! N \! \left( t=0 \right)$. $C$ is a positive, sufficiently large number, e.g., $C=50$. 
We choose $\epsce \! = \! 0.5$ which yields the correct $I_{\mathrm{p}}^{\mathrm{\left( 2 \right)}} \! = \! 2.0$. In order to reproduce $I_{\mathrm{p}}^{\mathrm{\left( 1 \right)}} \! = \! 0.904$ in the Schr\"odinger equation $\epsee \! = \! 0.329$ and in the TDDFT calculation $\epsee = 0.343$ are used (results do not change qualitatively for a different choice of $\epsce$ and $\epsee$).
To encompass the bound states, the parameter $a$ is set to $a \! = \! 6 $ throughout this work.
The Schr\"odinger and Kohn-Sham equations are solved numerically by a split-operator time propagator on a numerical grid (see, e.g., \cite{BauerKoval2006} and references therein).
 
For our model atom the diagonal of the two-electron density matrix is the pair density 
$\rho \left( x_{1},x_{2},t \right) = 2 \, \vert \psi \left( x_{1},x_{2},t \right) \vert^{2}$.
The exact density of the system
\begin{equation}
n \left( x, t \right)  = \int \mathrm{d}x' \, \rho \left( x,x',t \right)
\label{Formula Density from Psi}
\end{equation}
can in principle be calculated from the Kohn-Sham orbital of a TDDFT calculation with the correct exchange-correlation potential $\vxc$ as $n \left( x, t \right)  = 2\, \vert \phi \left( x,t \right) \vert^{2}$. Using $\vxc$ in the LK05-approximation yields approximate densities $n_{\mathrm{LK05}}$.
The time-dependent correlation function of the model atom is $ \gxc \left[ n \right] \left( x_{1},x_{2},t \right) = \rho \left( x_{1},x_{2},t \right) / n \left( x_{1},t \right) \, n \left( x_{2},t \right)$. 
The conditional probability to find at time $t$ an electron at $x_{2}$ in $\mathrm{d}x_{2}$ \emph{if} there is an electron at $x_{1}$ is given by $n \left( x_{2},t \right) \, \gxc \left( x_{1},x_{2},t \right)\, \mathrm{d}x_{2}$
Since the probability is an observable, the Runge-Gross theorem \cite{RungeGross1984} assures that 
$\gxc$
can in principle be expressed exactly as a functional of the density \cite{PetersilkaGross1999}. 
   
For a system with a two-electron wavefunction that can be expressed as a product of single-electron orbitals, $\gxc$ simplifies to  $\gxc\!=\!\gx\!=\!\frac{1}{2}$. The correlation contribution to the correlation function is thus defined as
\begin{equation}
\gc \left[ n \right] \left( x_{1},x_{2},t \right) = \frac{\rho \left( x_{1},x_{2},t \right)}{n \left( x_{1},t \right) \, n \left( x_{2},t \right)}-\frac{1}{2} \, .
\label{Formula gc}
\end{equation}
Due to (\ref{Formula Density from Psi}), $\gc$ fulfills the integration constraints
$\int \mathrm{d}x_{i} \, n\left( x_{i},t \right) \, \gc \left[ n \right] \left( x_{i},x_{j \neq i},t \right)= 0$ with $i,j \in \{1,2\}$.

A quantity of central importance to our work is the number of electrons bound to the Helium core at time $t$, 
\begin{equation}
N(t)=\int_{-a}^{+a} \mathrm{d}x \, n \left(x,t \right) \, ,
\label{Formula N from n}
\end{equation}
which also plays a central role in the exchange-correlation potential $\vxc$ in ionization processes, as in Eq.\,(\ref{Formula vhxcLK05}). 

\begin{figure}[!htb]
\centering
\includegraphics[angle=0, width=0.4\textwidth]{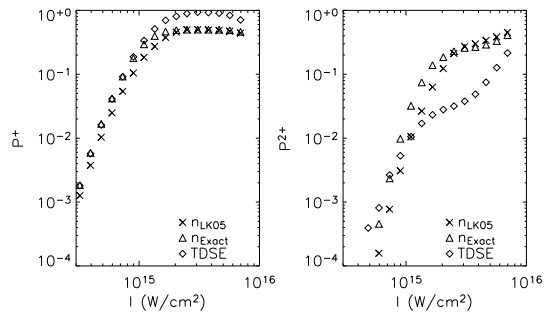}
\caption{Single and double ionization probabilities for exact and LK05 densities using $\Ic=0$ compared to the TDSE  solution for a $\lambda \! = \! 780\, \mathrm{nm}$
laser pulse.\label{Figure Ionization Ic=0}}
\end{figure}

In the two-electron space of the model atom we attribute areas to single ionization $P^{+}$ (either $\vert x_{1} \vert \! > \! a$ and $\vert x_{2} \vert \! \leq \! a$  or $\vert x_{1} \vert \! \leq \! a$ and $\vert x_{2} \vert \! > \! a$) and double ionization $P^{2+}$ ($\vert x_{1} \vert \! > \! a$ and $\vert x_{2} \vert \! > \! a$). Integration of $\vert \psi \left( x_{1},x_{2},T \right) \vert^{2}$ over these areas yields the ionization probabilities. This procedure to calculate ionization probabilities is well-established \cite{Bauer1997,LappasLeeuwen1998,DahlenLeeuwen2001} and leads to ionization probabilities depicted in Fig.\,\ref{Figure Ionization Ic=0}. Substituting the density and the correlation function for the pair density in the integrals and using $P^{0}+P^{+}+P^{2+}=1$ ($P^{0}$ being the probability for the atom to remain neutral) yields the exact ionization probabilities \cite{PetersilkaGross1999}
\begin{eqnarray}
P^{+} \left(t\right) & = & 2\, p\left( t \right) \left[ 1-p\left( t \right) \right] - \Ic \left( t \right) \, ,
\label{Formula P(He+) gc} \\
P^{2+} \left(t\right)& = & \left[ 1-p\left( t \right) \right]^{2}+\frac{1}{2} \, \Ic \left( t \right) \, ,
\label{Formula P(He++) gc}
\end{eqnarray}
where $p\left( t \right) = \frac{1}{2} \int_{-a}^{+a} \mathrm{d}x \, n \left(x,t \right)$ and the correlation integral is given by
\begin{equation}
\Ic \left( t \right)= \int_{-a}^{+a} \mathrm{d}x_{1} \int_{-a}^{+a} \mathrm{d}x_{2} \, n \left( x_{1} \right) \, n \left( x_{2} \right) \, \gc \left( x_{1},x_{2} \right) \, ,
\label{Formula Ic}
\end{equation}
with the density and the correlation function depending on time $t$ and $\gc=\gc \left[n \right]$.
   
It is important to note that Eqs.~(\ref{Formula P(He+) gc})-(\ref{Formula Ic}) assure that only the exact density in the range $-a \leq x \leq +a$ and the exact correlation function 
for 
$-a \leq x_{1},x_{2} \leq +a$ are required to calculate the exact ionization probabilities.

The functional dependence of $\gc$ on $n$ is not known and the calculation of the pair density $\rho$ at time $t=T$ is computationally demanding even for the one-dimensional Helium model atom. Therefore, it is necessary to find a suitable approximation for $\gc \left[ n \right] \left( x_{1}, x_{2}, T \right)$ to calculate $\Ic$.
Setting $\Ic \left( t \right) \! = \! 0$ corresponds to assuming the two-electron wavefunction to be a product of identical single-electron orbitals. This interpretation of the Kohn-Sham orbitals as electron orbitals has been widely used \cite{Bauer1997,LappasLeeuwen1998,PetersilkaGross1999,LeinKuemmel2005}. However, this approach leads to ionization probabilities which differ significantly from the exact probabilities even when the exact density (\ref{Formula Density from Psi}) is used \cite{LappasLeeuwen1998}, as shown in Fig.\,\ref{Figure Ionization Ic=0}. The LK05 approximation for $\vxc$ leads to probabilities which closely resemble the probabilities calculated from the exact densities.
       
\begin{figure}[!htb]
\centering
\includegraphics[angle=0, width=0.21\textwidth]{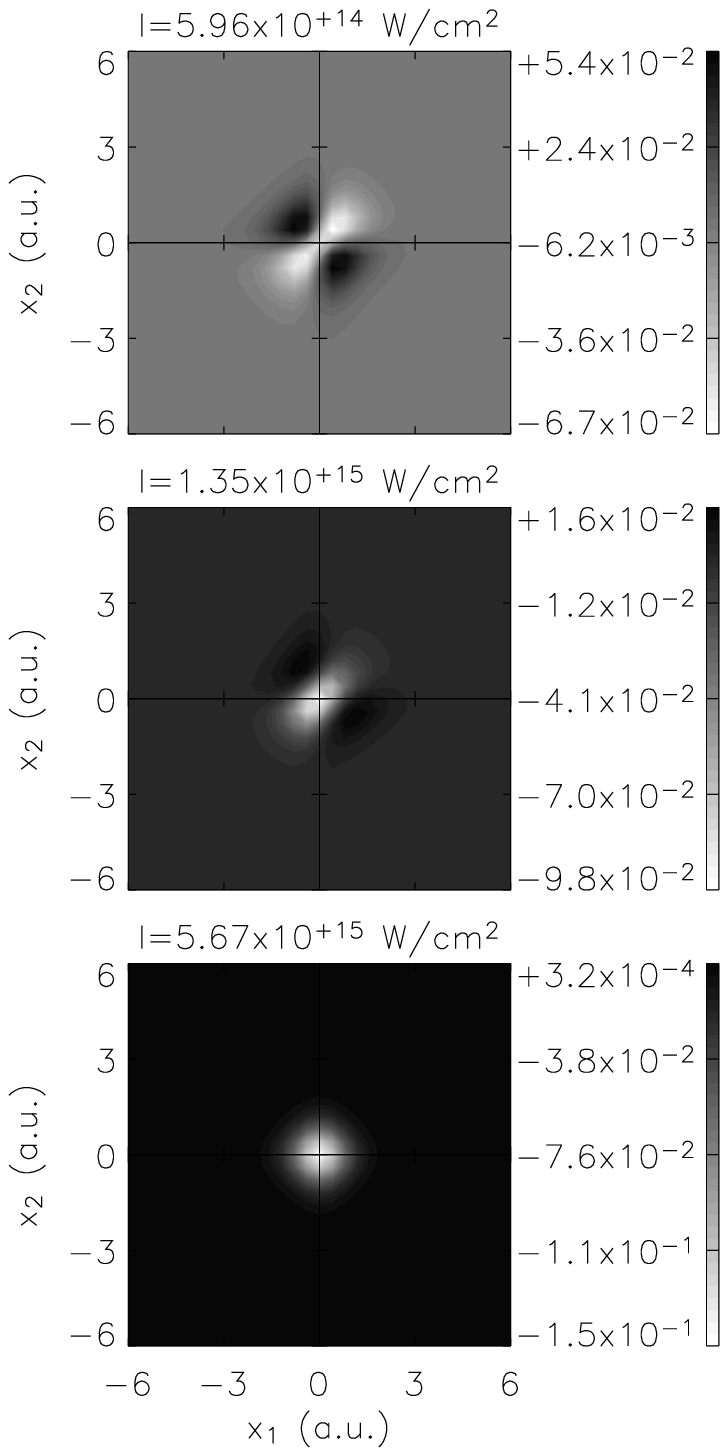}
\includegraphics[angle=0, width=0.21\textwidth]{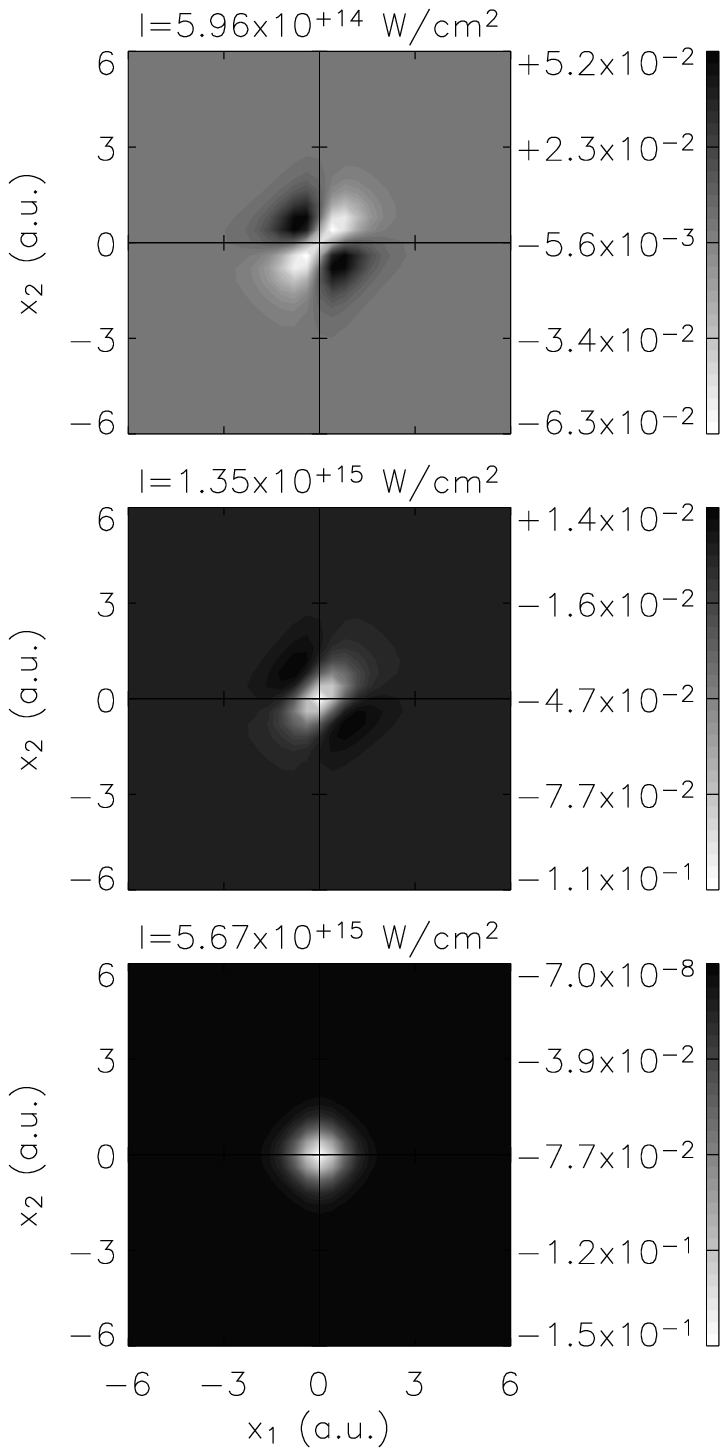}
\caption{Comparison of the integrand of the exact $\Ic \left( T \right)$ (left) and the adiabatic approximation $\IcA \left( T \right)$ (right) for different effective peak intensities of a $\lambda\!=\!780\, \mathrm{nm}$
laser pulse. \label{Figure Ic Contour}}
\end{figure}

The TDSE solution yields the exact pair density.
This allows us to calculate the exact $\gc$ and $\Ic$ as a reference for approximations of $\gc$. Figure \ref{Figure Ic Contour} reveals a structure 
of the integrand of $\Ic \left( T \right)$ which is remarkably simple. This is caused by the multiplication of $\gc$ by the density at $x_{1}$ and at $x_{2}$, which exhibits a maximum at the origin. For the 
other laser pulse
the integrand of $\Ic$ reveals a qualitatively similar structure.       
       
\begin{figure}[!htb]
\centering
\includegraphics[angle=0, width=0.21\textwidth]{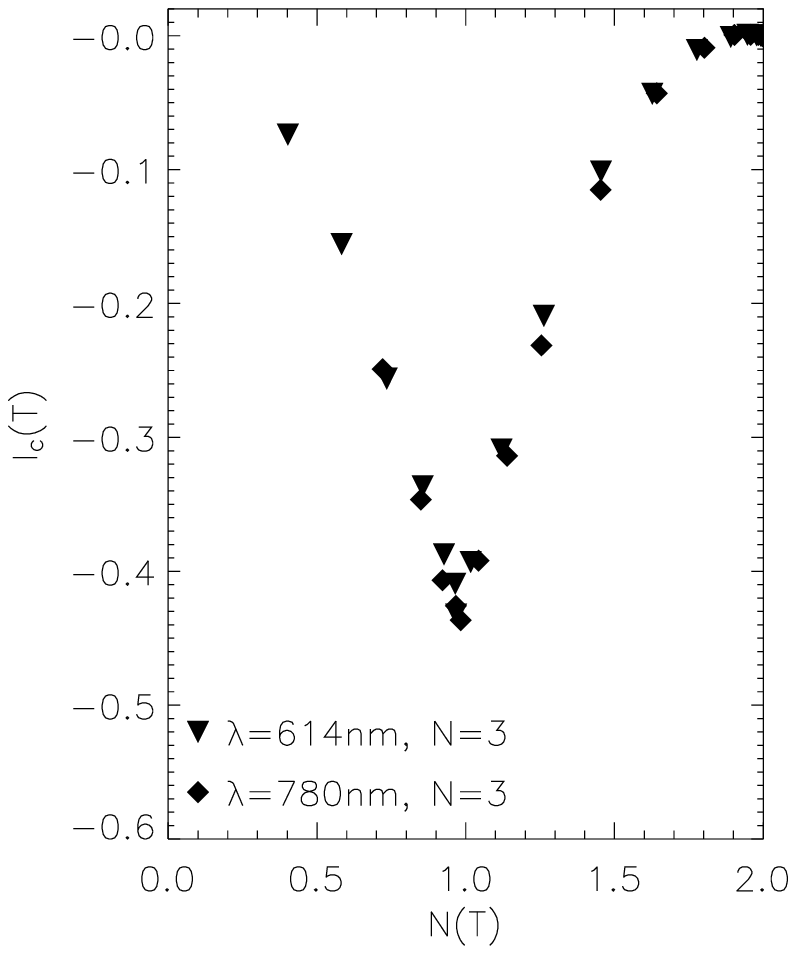}
\includegraphics[angle=0, width=0.21\textwidth]{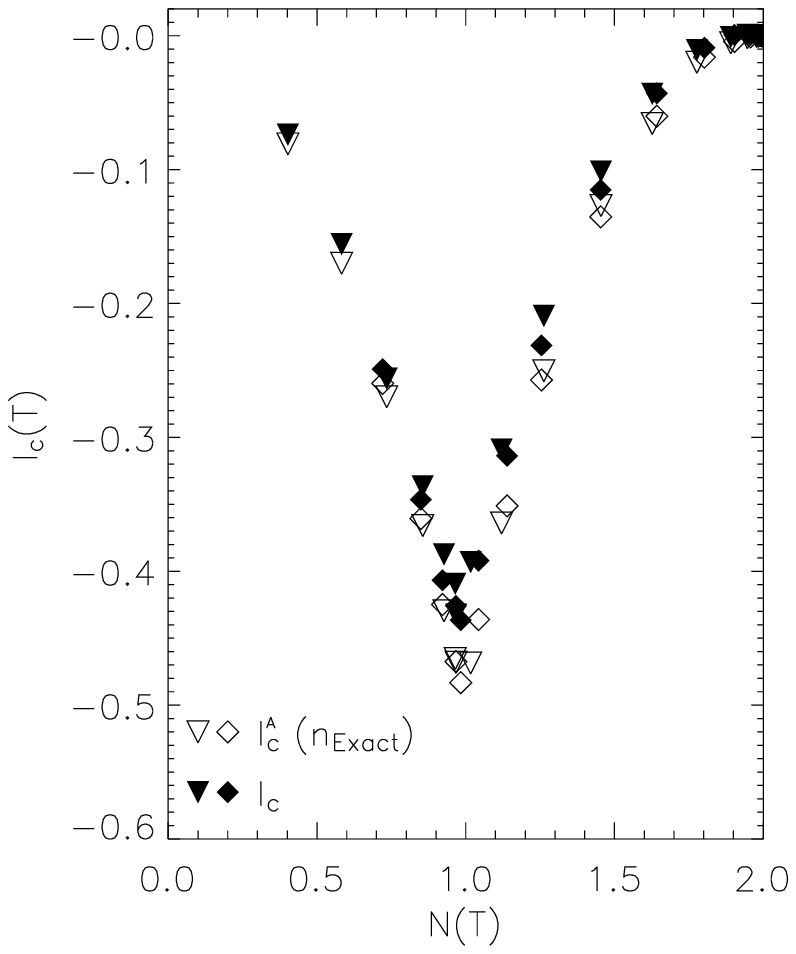}
\caption{Value of the exact $\Ic \left( T \right)$ as a function of the number of bound electrons (left).
Value of $\IcA \left( T \right)$ using exact densities compared to the exact $\Ic \left( T \right)$
(right). Results for both laser pulses are shown together with differently shaped symbols.
\label{Figure Value Ic}}
\end{figure}

The values of $\Ic \left( T \right)$ for the 
two
laser pulses are depicted in the left panel of Fig.\,\ref{Figure Value Ic} as a function of the number of bound electrons $N\! \left( T \right)$. $\Ic$ shows a quantitatively similar behavior for 
both
laser pulses: a minimum value for $\Ic$ is reached when the number of electrons is equal to one. This result points to the importance of the quantity $N\! \left( T \right)$ for the value of $\Ic \left( T \right)$.

We shall now introduce our adiabatic approximation for $\gc$. The starting point is our assumption that the system in the vicinity of the Helium core after the laser pulse closely resembles a groundstate system with a fractional number of electrons given by the number of bound electrons $N \left( T \right)$. We approximate numerator and denominator of the correlation function separately as linear combinations of the groundstate densities $n_{i}$ and groundstate pair densities $\rho_{i}$ of $i$-electron systems.
The density of the model Helium atom is expressed as \cite{LeinKuemmel2005}
\begin{equation}
\nA  =  \left\{ \begin{array}{ll}	\left( 1-N \right) \, n_{0} + N \, n_{1} & \ 0  \leq \! N \! \leq 1 \\
\left( 2-N \right) \, n_{1} + \left( N\! -1 \right) \, n_{2} & \ 1 \leq \! N\! \leq 2 \end{array} \right. \! ,
\label{Formula n Adiabatic}
\end{equation}
where $\nA = \nA \left(x ,t \right)$, $n_{i}=n_{i} \left( x \right)$ and $N=N\left( t \right)$ is given by (\ref{Formula N from n}).
Using $\nA \left( t \right)$ in the groundstate Kohn-Sham equation to calculate $\vc \left( t \right)$ and consequently employing $\vc \left( t \right)$ in the time-propagation has been shown to improve ionization probabilities compared to $\vc=0$ \cite{deWijnKuemmelLein2006II}.
Since $n\left( x \right) \mathrm{d}x$ is the probability to find an electron at $x$ in $\mathrm{d}x$, the density of the zero-electron system vanishes: $n_{0}=0$. Assuming the same adiabatic dependence of $\rho^{A} \left( t \right)$ on $N \! \left(t\right)$ the pair density is approximated as
\begin{eqnarray}
\rhoA  = \left\{ \begin{array}{ll}
\left( 1-N \right) \, \rho_{0} + N \, \rho_{1} & \ 0 \leq  \! N \! \leq 1 \\
\left( 2-N \right) \, \rho_{1} +\left( N\! -1 \right) \, \rho_{2} & \ 1 \leq \! N \! \leq 2 \end{array} \right. \! ,
\label{Formula rho Adiabatic}
\end{eqnarray}
where $\rhoA=\rhoA \left(x_{1},x_{2} ,t \right)$, $\rho_{i}=\rho_{i} \left(x_{1},x_{2} \right)$ and $N=N\left( t \right)$.
The pair density $\rho \left( x_{1},x_{2}\right)$ gives the joint probability to find an electron at $x_{1}$ in $\mathrm{d}x_{1}$ and an electron at $x_{2}$ in $\mathrm{d}x_{2}$. Corresponding to our consideration for the density, the pair densities of the zero- and one-electron systems are zero: $\rho_{0}=0$, $\rho_{1}=0$.
For our model atom, $n_{1} \left( x \right)$ is the groundstate density of $\mathrm{He^{+}}$. It is calculated from the solution of the respective one-dimensional stationary Schr\"odinger equation with 
$\epsce \! = \! 0.5$.
Consistently we use $n_{2} \left(x \right)= n \left( x,t=0 \right)$ where $n$ denotes the density utilized to calculate $N\! \left( t \right)$ in (\ref{Formula N from n}). Our adiabatic approximation of $\gc$ thus is
\begin{equation}
\gcA = \left\{ \begin{array}{ll}
-\frac{\textstyle 1}{\textstyle 2} & \ 0 \leq N \! \leq 1 \\
\frac{\textstyle \rho^{\mathrm{A}} \left(x_{1},x_{2},t \right)}{\textstyle n^{\mathrm{A}} \left(x_{1},t \right) \, n^{\mathrm{A}} \left(x_{2},t \right)}-\frac{\textstyle 1}{\textstyle 2} & \ 1 \leq N \! \leq 2
\end{array} \right. \! ,
\label{Formula g_c Adiabatic}
\end{equation}
with $\gcA = \gcA \left( x_{1},x_{2},t \right)$.
Although for $N(T)=1$ one has $\gcA=1/2$ which yields the unphysical value $P^{2+} \left(T\right)=0$, good results are already acquired for $N \left( T \right)$ only slightly smaller or larger than one, as is shown below.
	
In this form the approximation recovers the exact correlation function for the groundstate. This approximation can only be valid in the immediate vicinity of the Helium core. However, as pointed out above this is sufficient to calculate the correlation integral $\IcA \left( T \right)$. The integration constraints $\int \mathrm{d}x_{i} \, n\left( x_{i},t \right) \, \gc \left[ n \right] \left( x_{i},x_{j \neq i},t \right)= 0$ with $i,j \in \{1,2\}$ are not fulfilled since for $\vert x_{i} \vert > a$ we have $n \neq 0$ and $\gc \neq 0$ but 
set $\gcA = 0$. 
The same densities as in Eq.\,(\ref{Formula N from n}) are used in the integrand of $\IcA$.

The groundstate inputs needed to calculate $\gcA$ are accessible numerically also for more complex systems than Helium. Moreover, in the three-dimensional case the groundstate pair density can be calculated approximately via one of the well-known approximations for the groundstate correlation function \cite{Becke1988,PerdewWang1992}.
	
The right hand side of Fig.\,\ref{Figure Ic Contour} shows the integrand of $\IcA \left( T \right)$ using the exact densities. The qualitative agreement with the exact integrand of $\Ic \left( T \right)$ is good. Furthermore, the maximum and minimum values of the integrand are close to the exact values. This is a strong indication that the adiabatic approximation of $\gc$ captures the essential features which contribute to $\Ic$. In the right plot of Fig.\,\ref{Figure Value Ic} the adiabatic approximation $\IcA$ is compared to the exact values $\Ic$ for 
both
laser pulses as a function of $N \! \left( T \right)$. The adiabatic approximation exhibits a very similar dependence on $N$ as the exact values. However, for $N\left( T \right) \approx 1.0$ the values differ from the exact results. In our adiabatic approximation for $N\! \left( t \right) \leq 1$ all correlation effects vanish, as can be seen from (\ref{Formula g_c Adiabatic}). This might indicate the necessity to include memory effects in a refined approximation of $\rho$.	
\begin{figure}[!htb]
\centering
\includegraphics[angle=0, width=0.4\textwidth]{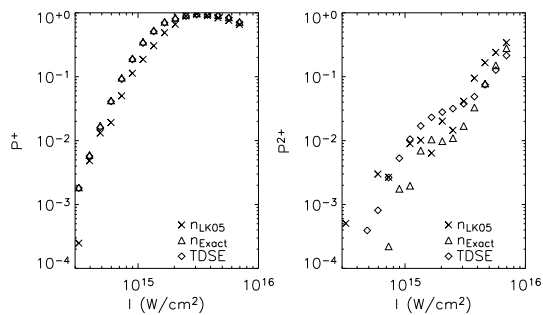}
\caption{Single and double ionization probabilities for exact and LK05 densities using the adiabatic approximation of the correlation integral $\IcA$ compared to the TDSE solution for a $\lambda \! = \! 780\, \mathrm{nm}$
laser pulse.\label{Figure Ionization IcA}}
\end{figure}
       
Figure \ref{Figure Ionization IcA} shows the ionization probabilities calculated from the exact and the LK05 densities using the approximated $\IcA \left( T \right)$ for the 
$\lambda \! = \! 780 \, \mathrm{nm}$
laser pulse. Single ionization probabilities agree very well with the exact probabilities from the TDSE for both exact and LK05 densities. The systematically too low values for $\Ic=0$ are corrected. Double ionization probabilities show a good agreement for high intensities. For low intensities the double ionization probabilities (and $\IcA$) are so small that already small deviations of $\IcA$ from the exact $\Ic$ lead to relatively large deviations in the logarithmic plot of the ionization probabilities and to minima and maxima in the double ionization probability. 
Since $\gcA=1/2$ at $N \left( T \right) \! = \! 1$, the adiabatically corrected probabilities for the two laser pulses examined show a minimum at intensities when the second derivative of the exact probabilities changes its sign at the onset of the ``knee''-structure (where $N \left( T \right) \simeq 1$).
Comparison with the values for $\Ic = 0$ (Fig.\,\ref{Figure Ionization Ic=0}) evidences a significant improvement of the TDDFT ionization probabilities using our approximation of the correlation function (\ref{Formula g_c Adiabatic}).

In summary, to acquire the exact, multiple ionization probabilities of atoms in strong laser fields from time-dependent density functional calculations, knowledge of the correlation integral $\Ic \left( T \right)$ is vital. For a one-dimensional Helium model atom, an adiabatic approximation of the correlation function $\gc$ close to the Helium core yields an approximation of $\Ic$ which agrees qualitatively and quantitatively well with the exact $\Ic$. The exact single ionization probabilities and the ``knee''-structure in the double ionization probability are reproduced. The (pair) density of Helium after the laser pulse can be approximated by a linear combination of atomic and ionic groundstate (pair) densities 
using the fractional number of bound electrons. This method opens the possibility to apply the adiabatic approximation in three dimensions and to the ionization of more complex atoms.

We would like to thank Manfred Lein for stimulating discussions and the Deutsche Forschungsgemeinschaft for supporting this work.


\end{document}